\documentclass{PoS}
\usepackage{graphicx}
\PoS{PoS(LAT2005)225}
\title{Non-perturbatively determined relativistic heavy quark action}
\ShortTitle{NP determined RHQ action}
\author{\speaker{Huey-Wen~Lin} and Norman~H.~Christ\\
Columbia University, USA\\
E-mail: \email{hwlin@phys.columbia.edu},
\email{nhc@phys.columbia.edu}}
\abstract{Preliminary results are presented in a step scaling
determination of the coefficients in the relativistic heavy quark
action. By matching finite volume, heavy-heavy and heavy-light
meson masses, we attempt to determine the four parameters ($m$,
$\zeta$, $c_B$ and $c_E$) in the on-shell-improved, heavy quark
action. In this report we carry out one step in this program by
matching two physically equivalent systems.  The first is a fully
relativistic calculation using a $24^3\times 48$, $1/a=5.4$ GeV
lattice with both the heavy and light quarks treated as domain
wall fermions.  The second calculation uses at $16^3\times 32$,
$1/a=3.6$ GeV lattice, a domain-wall light quark and a heavy quark
computed with the relativistic heavy quark action.  The four
parameters in this heavy quark effective action are then varied to
reproduce the mass spectrum from the first calculation.  These
calculations are carried out in the quenched approximation for a
heavy quark mass approximately that of the charmed quark.}
\FullConference{XXIIIrd International Symposium on
Lattice Field Theory\\
 25-30 July 2005\\
Trinity College, Dublin, Ireland\\}
\begin{document}
\bibliographystyle{JHEP}

\section{Introduction}
Heavy quark physics plays an important role in determining the
basic parameters of the Standard Model~\cite{Kronfeld:2003sd,
Wingate:2004xa} and lattice QCD provides a first principles method
for determining these parameters.  However, to treat a charm or
bottom quark using presently accessible lattice spacings requires
the use of an effective theory, one which is not accurate for
energy scales on the order of the charm or bottom mass but which
will accurately describe the physics of charmed or bottom states
at the energy of scale of interest, the $\Lambda_{QCD}$  scale of
non-perturbative QCD.  Here we will study the relativistic heavy
quark effective theory of the Fermilab~\cite{El-Khadra:1997mp} and
Tsukuba~\cite{Aoki:2001ra} groups.

This relativistic heavy quark action can be written down as:

\vskip -4mm
\begin{eqnarray}
S & = & \sum_x \overline{\psi}(x)[m_0
+ \gamma_0 D_0 + {\zeta} \vec{\gamma} \cdot \vec{D}%
- {r_t} D_0^2 - {r_s} \sum_i{D_i}^2 \nonumber\\
&-& \sum_i \frac{i}{2}\sigma_{0i}F_{0i} + \sum_{i,j}\frac{i}{2}
c_{B} \sigma_{ij}F_{ij} + \xi \{D_0,D_i\}\sigma_{0,i}]\psi(x)
\label{eq:rhq_action}
\end{eqnarray}
\vskip -2mm
where \vskip -10mm
\begin{eqnarray}
D_{\mu}\psi(x) & = & \frac{1}{2}[U_{\mu}(x)\psi(x +\hat{\mu}) -
U_{\mu}^{\dagger}(x-\hat{\mu})\psi(x-\hat{\mu})]\\
D_{\mu}^2\psi(x)& = &\frac{1}{2}[U_{\mu}(x)\psi(x+\hat{\mu})
+U_{\mu}^{\dagger}(x-\hat{\mu})\psi(x-\hat{\mu})-2\psi(x)]\\
F_{\mu, \nu}\psi(x)&=& \frac{1}{8 i}\sum_{s,s'=\pm 1}
ss'[U_{s\mu}(x)U_{s'\nu}(x+s\hat\mu)\nonumber\\
&& \times
U_{-s\mu}(x+s\hat\mu+s'\hat\nu)U_{-s'\nu}(x+s'\hat\nu)
-h.c.\;]\psi(x)
\end{eqnarray}

As discussed in Ref.~\cite{El-Khadra:1997mp}, this action can be
used to compute amplitudes which involve heavy quarks carrying
spatial momenta of $O(p)$ which will be accurate to order
$F(ma)(pa)^2$.  Potentially large errors of order $(ma)^n$, where
$m$ is the heavy quark mass, can be removed by a proper choice of
the seven $ma$-dependent parameters, $m$, $\zeta$, $r_t$, $r_s$,
$c_E$, $c_B$ and $\xi$.  If the coefficient functions $F(ma)$ are
bounded, then these errors vanish uniformly in the $a\rightarrow
0$ limit.

As pointed out in Ref.~\cite{Aoki:2001ra}, the equations of motion
can be used to justify setting $r_t=1$ and $\xi=0$.  With this
choice all on-shell Greens functions will still take their
continuum form with errors of order $F(ma)(pa)^2$.  As explained
in Ref.~\cite{El-Khadra:1997mp}, a final field transformation can
be made to justify setting the parameter $r_s=\zeta$. Such a
transformation will not change particle masses but will result in
on-shell fermion ({\it e.g.} nucleon) propagators which do not
show the standard continuum form.  (Note this field transformation
is actually performed on the effective continuum action and
establishes a relation between the low-energy, on-shell Greens
functions of the two theories only.)  In the work presented here
we will make the Fermilab choice, $r_s=\zeta$ and determine the
remaining four parameters, $m$, $c_E$, $c_B$ and $\zeta$ using
non-perturbative methods that involve only the particle spectrum.
While not discussed here, the effects of this choice for $r_s$
could be determined from the spinor structure of the nucleon
propagator and then removed.
%

\section{Step-scaling strategy}
\label{SecProposal} 
\begin{figure}[hbt]
\centering
\includegraphics[width=0.8\columnwidth]{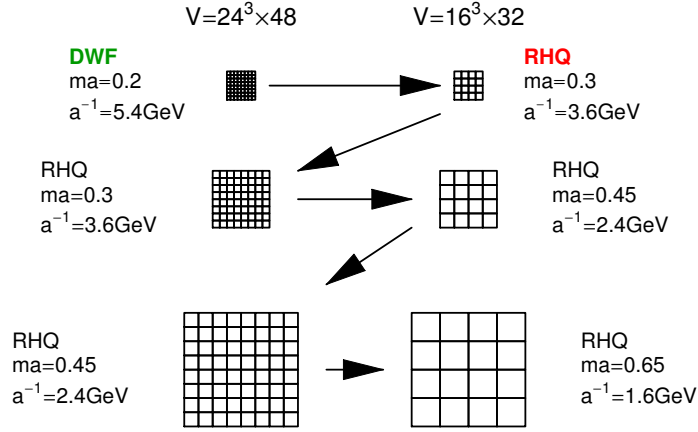}
\vskip -3mm
\caption{\small\small
A schematic of the step scaling technique we use to determine
the $m$, $c_B$,$c_E$ and $\zeta$, appropriate for charm
quark physics, from non-perturbative, $O(a)$-improved, light quark
calculations at $1/a=5.4$ GeV.} \label{step_scaling} \vskip -3mm
\end{figure}

We propose to determine these four coefficients
in the RHQ action by matching the finite-volume, heavy-heavy and
heavy-light spectrum with that determined in an accurate, small-$ma$
calculation, a strategy similar to that employed for the static
approximation~\cite{Heitger:2003nj}.  By performing a series of such
comparisons, see Fig.~\ref{step_scaling}, we can move from an accurate,
$ma \ll 1$ calculation in small volume, to a final action at a coarser
lattice spacing, practical for large-volume  calculations.

Figure~\ref{step_scaling} shows the steps that we have begun
to carry out.  Using the quenched approximation for simplicity,
we match physical quantities calculated on the two finest lattices,
which have a fixed physical volume of $(0.9fm)^3$.  For the $1/a=5.4$ GeV
lattice we use domain wall fermions (DWF), involving the single parameter,
$m \approx 1.08$ GeV, chosen to approximate the charm quark mass.  We
then adjust the coefficients of the RHQ action on the $1/a=3.6$
lattice until the spectra of the two calculations agree.  Second,
we expand the volume to $(1.35fm)^3$, keeping all coefficients
fixed, and then match with a fourth calculation with 3/2 larger
lattice spacing. Here, we demonstrate the practical
implementation of our approach through stage one: matching the
DWF 5.4 GeV (fine lattice) spectrum calculated that on a 3.6 GeV
(coarse lattice) using the RHQ action.

In the matching step we compare the pseudo-scalar(PS), vector(V),
scalar(S), and pseudo-vector(PV) meson masses for heavy-heavy(hh)
states and PS and V masses for heavy-light(hl) states.  We also
require the equality of the masses $m_1$ and $m_2$ in the hh-PS
dispersion relation $E(p^2)^{hh}_{PS} = m_1^2 + (m_1/m_2)p^2$.  For
both the heavy quark on the fine lattice and the light quark at
both lattice spacings we use the DWF action, see Ref.~\cite{Blum:2000kn}
for the method used here and further references.  In each case we
used a fifth dimensional extent, $L_s=12$ and a domain wall ``height'',
$M_5=1.5$, ensuring that there are no unphysical propagating states
for the quark masses used.

As a starting point for our $1/a=3.6$ GeV calculation, we used the
one-loop perturbative coefficients for the five-parameter Tsukuba
action~\cite{Aoki:2003dg}.  These were translated into the four
parameters of the Fermilab RHQ action by performing the $O(a)$
tree-level field transformation:
\begin{equation}
\psi_T       = [1+ \delta \vec{\gamma}\vec{D}]\psi_F, \quad%
\overline{\psi}_T = \overline{\psi}_F[1 - \delta \vec{\gamma}
\stackrel{\leftarrow}{D}], \quad %
\delta = \frac{- r_s - \zeta}{2(m_0 + \zeta)}.
\end{equation}

\section{Simulation}
\label{SecSimulation} 
\begin{figure}[hbt]
\begin{center}
\vskip 6mm
\includegraphics[width=0.5\columnwidth]{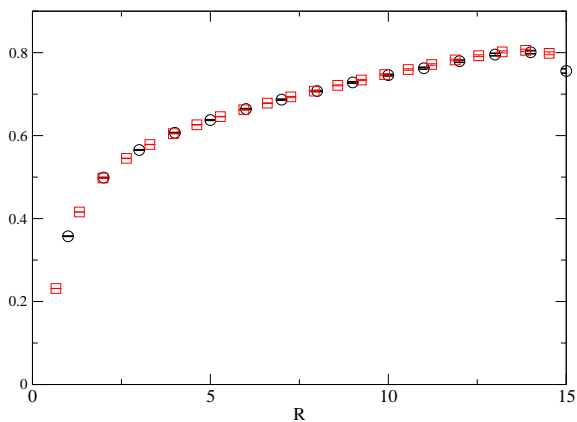}
\end{center}
\vskip -6mm
\caption{\small The squares are the static quark potential from
the $\beta$ = 6.638 lattice scaled to match the lattice spacing at
$\beta$ = 6.351.   The circles are the $\beta$ = 6.351 potential.} \label{static_ratio} \vskip -3mm
\end{figure}

We used the Wilson gauge action since its
relation between coupling and lattice spacing has been carefully studied~\cite{Necco:2001xg}.
However, as a final check, we examined
the static quark potential on our two ensembles.  We first determined
the lattice spacing directly from the two standards scales, $r_0$ and
$r_I$, obtaining results distorted by our small-volumes.  We also
obtained the ratio of lattice spacings directly from the following
relation:
\begin{eqnarray}
V_1(n_1)= V_2(n_2/\lambda)/\lambda + C
\label{eq:pot_scaling}
\end{eqnarray}
where $\lambda(= {a_2}/{a_1})$ is the lattice spacing ratio between
two lattices.  The potential $V_2(r)$, from the fine lattice, is fit
to a standard phenomenological form and then $\lambda$ determined so
that this fit, scaled as in Eq.~\ref{eq:pot_scaling} matches that on
the coarse lattice, giving the expected ratio $a_1/a_2=1.51(2)$.  Figure.~\ref{static_ratio} shows how accurately these scaled static
quark potentials agreed.

We employed a ``smeared'', Coulomb gauge-fixed, heavy quark wavefunction
source using orthogonal hydrogen ground and first-excited wavefunctions.
Table~\ref{tab:coarsePar} lists the 22 RHQ parameter sets that we have
run on the coarse lattices.  The first few sets of data were chosen close
to the 1-loop coefficients described above and the latter ones adjusted as
we further explored the parameter space.
\begin{center}
\vskip -6mm
\begin{table}[hbt]
\caption{\label{tab:coarsePar} The details of the parameters we
have run for the work reported here. }
{\small
\begin{minipage}{7cm}
\begin{tabular}{c|c|c|c|c}
Label & $m_0$       &    $c_B$   &     $c_E$  &    $\zeta$  \\
\hline %
   1 &  0.00      & 1.552 & 1.458 &  1.013 \\
   2 &  0.07      & 1.547 & 1.424 &  1.001 \\
   3 &  0.0426    & 1.550 & 1.438 &  1.007 \\
   4 &  0.0426    & 1.550 & 1.438 &  1.100     \\
   5 &  0.0426    & 1.550 & 1.438 &  0.900     \\
   6 &  0.0330    & 1.609 & 1.538 &  1.044 \\
   7  &  0.0230   & 1.609 & 1.438 &  1.044  \\
   8  &  0.0230   & 1.609 & 1.538 &  1.044  \\
   9  &  0.0426   & 1.609 & 1.438 &  1.044  \\
   10  &  0.0426  & 1.550 & 1.438 &  1.007  \\
   11  &  0.0     & 1.552 & 1.438 &  1.013  \\
\end{tabular}
\end{minipage}
\hspace{0.1cm}
\begin{minipage}{7cm}
\begin{tabular}{c|c|c|c|c}
Label & $m_0$       &    $c_B$   &     $c_E$  &    $\zeta$  \\
\hline %
   12  &  0.0328  & 1.511 & 1.438 &  1.036  \\
   13  &  0.0230  & 1.511 & 1.438 &  1.036  \\
   14  &  0.0328  &  1.511 &  1.538  &  1.036  \\
   15  &  -0.01   &  1.700 &  1.574  &  1.022  \\
   16  &  0.00371 &  1.709 &  1.577  &  1.023  \\
   17  &  0.0138  &  1.715 &  1.579  &  1.025  \\
   18  &  0.02    &  1.719 &  1.580  &  1.026  \\
   19  &  0.03    &  1.725 &  1.582  &  1.027  \\
   20  &  0.08    &  1.707 &  1.576  &  1.023  \\
   21  &  0.09    &  1.713 &  1.578  &  1.025  \\
   22  &  0.1     &  1.719 &  1.580  &  1.026  \\
\end{tabular}
\end{minipage}
}
\end{table}
\end{center}

\vskip -0.5 in
\section{Analysis and result}
\label{SecAnalysis}
We next use the above parameter sets to determine the dependence
of the measured quantities on the four RHQ action parameters we
wish to determine.  We attempt to work within a parameter range
where we can use the linear relation:
\begin{equation}
Y_{coarse}^i = A + J \cdot X_{RHQ}^i
\label{eq:linear}
\end{equation}
Here $A$ is a $d$-dimensional vector and $J$ an $d \times 4$ matrix.
The vector $X_{RHQ}^i$ is formed of the four RHQ parameters
$\{(X_{RHQ})_a^i\}_{1 \le a \le 4} = \{m_0^i, c_B^i, c_E^i, \zeta^i\}$
while the $d$-dimensional vector $Y_{coarse}^i$ is formed from the
spectral quantities measured on the coarse lattice.  For $d=6$
$\{(Y_{coarse}^i)_a\}_{1 \le a \le 6} = \{\frac{1}{4}(M_{PS}+3M_V)^{hh,i},
(M_{PS}-M_V)^{hh,i},(M_{PV}-M_S)^{hh,i},(m_1/m_2)^{hh,i},
\frac{1}{4}(M_{PS}+3M_V)^{hl,i},(M_{PS}-M_V)^{hh,i}\}$.  The index $i$
labels the parameter set and varies between 1 and 22.

Equation~\ref{eq:linear} can be solved directly for $A$ and $J$ if we
use five parameter sets.  More parameter sets can be used if we minimize
a weighted sum of the deviations between the left- and right-hand sides
of Eq.~\ref{eq:linear}.  We can then use the resulting values for $J$ and
$A$, to solve for the parameters $X_{RHQ}$ that yield meson masses as
close as possible to those on the fine lattice (denoted as $Y_{fine}$
with error $\sigma_{fine}$) by minimizing the following quantity with
respect to $X_{RHQ}$:
\begin{eqnarray}
\chi_{fine}^2 = \sum_{n=1}^{d} \frac{|(J\cdot X_{RHQ} + A - Y_{fine})_n
|^2}{\sigma_{fine,n}^2}.
\label{eq:chi_sq}
\end{eqnarray}

We divided our data into three categories to explore the sensitivity
of the resulting coefficients to our choice of masses being matched:
A. heavy-heavy system only, including spin-orbit splitting; B.
heavy-heavy (no SO splitting) and heavy-light systems; and C. all of
the above measurements.  The first five rows of Table~\ref{tab:result}
demonstrate the stability of our results for the least constrained
choice of set A.  As we incrementally removed the parameter sets that
contribute the most to the chi-squared in Eq.~\ref{eq:chi_sq}, the
matching RHQ parameters remain consistent.
\begin{table}[hbt]
\caption{\label{tab:result} A check on the stability of the
analysis using the example of data set A and results for B and C.
Note, the errors on $m_0$ should be viewed as ~20\% of the
$\approx 0.3$ renormalized mass.}
\begin{center}
\begin{tabular}{l|c|r|r|r|r}
data sets &$\chi^2/d.o.f.$
                   & \multicolumn{1}{c}{$m_0$}
                                 & \multicolumn{1}{c}{$c_B$}
                                            & \multicolumn{1}{c}{$c_E$}
                                                      &\multicolumn{1}{c}{$\zeta$}   \\
\hline
A(all)      & 512/22 & 0.035(61)     & 1.725(10) & 1.3(5)  & 1.036(17) \\
(drop 15)   & 123/21 & 0.034(60)     & 1.726(10) & 1.3(5)  & 1.036(17) \\
(drop 4,15) & 57/20  & 0.027(66)     & 1.71(9)   & 1.3(4)  & 1.039(18) \\
(drop 3, 4, 15)
            & 41/19  & 0.026(67)     & 1.71(9)   & 1.3(4)  & 1.039(18) \\
(drop  2, 3, 4, 15)
            & 29/18  & 0.026(65)     & 1.70(9)   & 1.3(4)  & 1.040(18) \\
\hline
B           & 128/21 & 0.03(21)      & 1.75(18)  & 1.3(17) & 1.04(4)   \\
\hline %
C           & 42/19  & 0.011(61)     & 1.73(1)   & 1.2(4)  & 1.043(17) \\
\end{tabular}
\end{center}
\end{table}
As Table~\ref{tab:result} suggests, we have very consistent results
for different choices of parameter sets and meson masses to be
matched.  The results from parameter set B, which excludes the
spin-orbit splitting, gives $c_E$ and $m_0$ with enormous errors.
On the other hand, the matching coefficients determined from sets
A and C look more promising.


\section{Outlook and Conclusion}
\label{SecConclusion}
\vskip -1mm
It appears practical to determine $m_{0}$,
$c_B$, $c_E$ and $\zeta$ from finite-volume, non-perturbative
matching. More statistics and better parameter coverage should
reduce the matching errors to a few percent. Carrying out
the second matching step in Fig.~\ref{step_scaling} will
permit us to perform a RHQ, charm physics calculation with
$1/a=2.4$ GeV and a $(2 {\rm fm})^3$ volume.  While we have
reported quenched results here, it should be emphasized that this
approach can be extended to full QCD without excessive
computational cost.  Since the RHQ parameters being determined
are short-distance quantities, they depend on both the lattice
volume and sea quark masses only through $O(a^2)$ errors.  Hence
the dynamical quark masses need not be physical but must only
obey $m_{\rm sea} \ll 1/a$.  Thus, we need only require that
$N_f=3$ and that $m_{sea}$/$\Lambda_{QCD}$ are equal for each
pair of systems being matched.


\section*{Acknowledgments}
\vskip -1mm We acknowledge helpful discussions with Tanmoy
Bhattacharya, Peter Boyle,  Paul Mackenzie and members of the RBC
collaboration. In addition, we thank Peter Boyle, Dong Chen, Mike
Clark, Saul Cohen, Calin Cristian, Zhihua Dong, Alan Gara, Andrew
Jackson, Balint Joo, Chulwoo Jung, Richard Kenway, Changhoan Kim,
Ludmila Levkova, Xiaodong Liao, Guofeng Liu, Robert Mawhinney
Shigemi Ohta, Konstantin Petrov, Tilo Wettig and Azusa Yamaguchi
for developing with us the QCDOC machine and its software. This
development and the resulting computer equipment used in this
calculation were funded by the U.S. DOE grant DE-FG02-92ER40699,
PPARC JIF grant PPA/J/S/1998/00756 and by RIKEN. This work was
supported by DOE grant DE-FG02-92ER40699 and we thank RIKEN,
Brookhaven National Laboratory and the U.S. Department of Energy
for providing the facilities essential for the completion of this
work.

\bibliography{Lat05Pro05}

\end{document}